\documentclass[useAMS,usenatbib]{mn2e}
\input{psfig}
\title[Possible radio supernova in NGC\,3310]
{A possible radio supernova in the outer part of NGC\,3310}
\author[]
{M. K. Argo\thanks{mkargo@jb.man.ac.uk}, T. W. B. Muxlow, A. Pedlar, R. J. Beswick, M. Strong\\
University of Manchester, Jodrell~Bank Observatory, Macclesfield, Cheshire SK11~9DL\\}

\date{12-03-04}
\def\ltsim  {\ifmmode\stackrel{<}{_{\sim}}\else$\stackrel{<}{_{\sim}}$\fi}
\def\gtsim  {\ifmmode\stackrel{>}{_{\sim}}\else$\stackrel{>}{_{\sim}}$\fi}
\def\farcs  {\hbox{$.\!\!^{\prime\prime}$}}
\def\kms    {\ifmmode {\rm km\,s}^{-1} \else km\,s$^{-1}$\fi}

\newcommand{\src}{J103851+532927 }
\begin{document}
\maketitle

\begin{abstract} 

{\large As part of an on-going radio supernova monitoring
program, we have discovered a variable, compact steep spectrum radio
source $\sim$65 arcsec ($\sim$4\,kpc) from the centre of the starburst
galaxy NGC\,3310. If the source is at the distance of NGC\,3310, then its 5\,GHz luminosity is
$\sim3 \times 10^{19}$\,W\,Hz$^{-1}$. The source luminosity, together with its variability 
characteristics, compact structure ($<$17\,mas) and its association with a group of H{\sc ii}
regions, leads us to propose that it is a previously uncatalogued type II radio supernova.  A search of 
archival data also shows an associated X-ray source with a luminosity similar to known radio supernova.

}
\end{abstract}


\section{Introduction}

The starformation rate (SFR) of a galaxy is an important parameter used
over a wide range of astrophysics.  A number of methods (e.g. Cram et al
1998) can be used to estimate this parameter using UV, optical, FIR and
radio observations. However the UV and optical methods are particularly
difficult in starbursts because of the large extinction corrections
required. The radio and infra-red methods seem to give reasonably
consistent results, resulting in the well known correlation between radio
and FIR luminosities. However it is important to note that neither the
radio nor FIR methods are completely understood theoretically.

In principle, an independent estimate of the SFR can be derived from the
supernova rate of a galaxy. If all stars with masses greater than 8\,M$_\odot$
become supernovae via core collapse then, if a reasonable initial mass
function is assumed, the SFR can be determined as a simple function of
supernova rate. This method is particularly applicable to regions such as
starbursts with high predicted supernova rates. However in starbursts, the
high levels of extinction due to the dust intrinsic to the starburst (e.g.
$\sim30$ magnitudes in M82 - Mattila \& Meikle 2001), can severely 
compromise optical searches for supernovae even in relatively nearby starburst
galaxies.  In contrast, the radio emission from supernovae will not be significantly
affected by extinction and hence we have begun a program of monitoring a
number of nearby starbursts using MERLIN (Multi-Element Radio Linked 
Interferometer Network) and the VLA (Very Large Array), primarily to detect
new radio supernovae, but also to monitor the flux density of existing
remnants.

The first monitoring session took place in November 2003 using the VLA in B 
configuration.  In the analysis of the radio emission from NGC\,3310, 
a relatively strong radio source was found in the outer part of the galaxy. 
A number of possible explanations for this source are considered, but the most likely 
conclusion is that it is a radio supernova.

Throughout this paper the distance to NGC\,3310 is assumed to be 13.3\,Mpc 
(H$_0$\,=\,75\,\kms\,Mpc$^{-1}$, Kregel \& Sancisi 2001) at which an angle of 1 arcsec
corresponds to a linear size of 64\,pc.

\section{Observations}

\begin{figure}
\psfig{figure=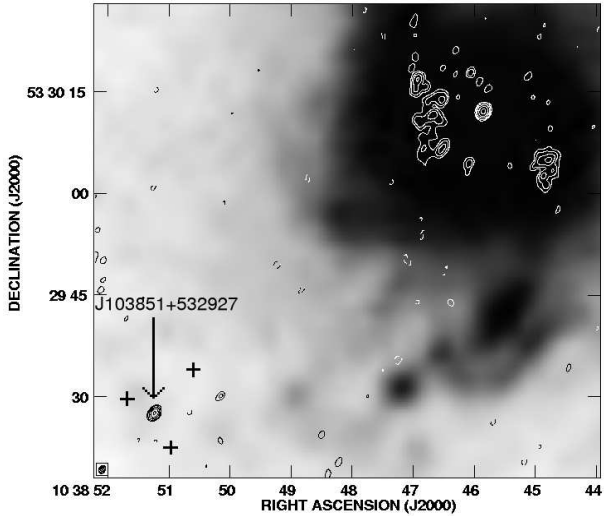,width=8cm}
\caption{\label{imagec}November 2003 4.885\,GHz image of NGC\,3310 (contours) overlaid on 
a Digitised Sky Survey optical image (grey scale).  Contours are 
plotted at -1, 1, 2, 4, 8, 16 and 32 $\times$ 95\,$\mu$Jy and the angular 
resolution is 1.24$\times$0.91 arcsec. The crosses mark the positions of three H{\sc 
ii} regions reported by Hodge \& Kennicutt (1983).}
\end{figure}

As part of the monitoring program the VLA was used in B configuration to observe 
NGC\,3310, along with 9 other starburst galaxies, at 5, 8.4 and 15\,GHz in November 2003.  
The observations were flux calibrated using observations of 3C286 and phase calibrated 
using the nearby calibrator 1035+564. Each data-set was imaged using standard techniques and
the image obtained at 5\,GHz, with an angular resolution of 1.24$\times$0.91 arcsec, is 
shown in Fig.~\ref{imagec}.
In this image, a number of
compact sources can be seen within the central 15 arcsec of NGC\,3310, in good
agreement with previous observations (e.g. Duric et al 1986, Mulder, van Driel \& Braine 
1995), however a relatively strong source (1.57\,mJy at 4.885\,GHz) can be seen $\sim$65\,arcsec 
to the SE.  This source - hereafter referred to as \src - is the subject of this 
paper and its position and flux densities are given in Table~\ref{fluxtab}.

In December 2003 through to January 2004 \src was also observed using MERLIN (Thomasson 1986) at 
5\,GHz. The phase and flux density were calibrated using 1041+536 and 
3C286 respectively. Only 4 antennas were available although, as 
this included the Cambridge antenna (as well as the Mk II, Knockin and Darnhall telescopes) 
it was possible to image this source at maximum 
angular resolution ($\sim$35\,mas at 5\,GHz).  A number of imaging runs on the source 
were undertaken and several  independent images were produced. 
The flux density, size and position of the source remained consistent for all these 
images and the final parameters are given in Table~\ref{fluxtab}.
All these  images showed the source to be unresolved and unpolarised ($<$2\,per\,cent).  
The position of the source, measured relative to the phase calibrator 
1041+536 (RA=10$^h$44$^m$10{\hbox{$.\!\!^{s}$}}6716
 Dec=+53$^o$22$^{\prime}$20\farcs522), is  10$^h$38$^m$51{\hbox{$.\!\!^{s}$}}24599
+53$^o$29$^{\prime}$27\farcs508  with an error of $\sim10$\,mas.

\begin{figure}
\psfig{figure=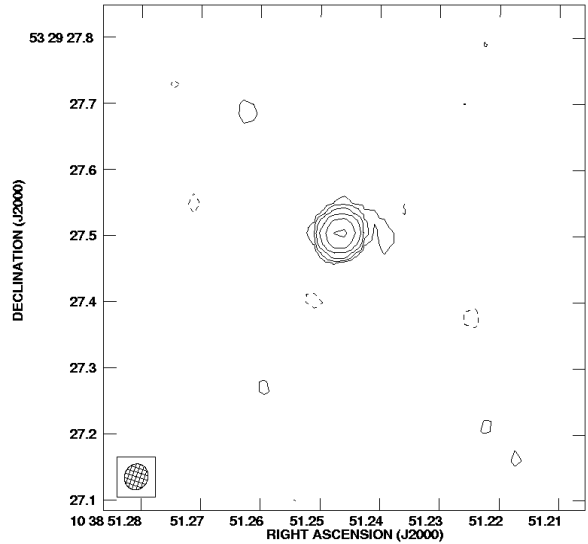,width=8cm}
\caption{\label{merlin}December 2003 MERLIN map of \src at 4.994 GHz.  Contours are 
plotted at -1, 1, 2, 4, 8 and 16 $\times$ 60\,$\mu$Jy.}
\end{figure}

A number of archival observations from both MERLIN and the VLA have also been 
imaged.  In all cases standard reduction techniques were used and the flux 
densities calibrated relative to 3C286.  The results of these archival 
observations are also given in Table~\ref{fluxtab}.

\begin{table*}
\begin{tabular}{cccccc}
Freq	& Epoch		& Flux ($\pm$3$\sigma$)	& \multicolumn{2}{c}{Position (J2000)}	& Telescope (ref)		\\
(GHz)	&		& mJy			& RA (10$^h$ 38$^m$)		& Dec (+53$^o$ 29$^{\prime}$)	&	 \\
\hline
1.465	& 1982.04     	& $<$1 			& -		& -	& VLA C (1)	\\ 
1.465	& 1986.21	& 0.79$\pm$0.23		& 51{\hbox{$.\!\!^{s}$}}230	& 27\farcs66	& VLA A (2)	\\
1.490	& 1986.62	& 1.62$\pm0.65$		& 51.235	& 27.83	& VLA B (2)	\\
1.4	& 1987.04	& $<$1			& -		& - 	& WSRT (3)	\\
1.4	& 1997.37	& 7.83$\pm$0.73		& 51.260	& 27.58	& VLA B (4)	\\
1.4	& 1997.54	& $\sim$7 		& 51.224	& 27.38	& WSRT (5)	\\
1.658	& 1999.91	& 5.54$\pm$0.34		& 51.245	& 27.50	& MERLIN (6)	\\
\hline
4.860	& 1986.62	& 0.25$\pm$0.17		& 51.228	& 27.55	& VLA B (2)	\\
4.994	& 2000.22	& 1.88$\pm$0.45		& 51.247		& 27.61	& MERLIN (7)	\\
4.885	& 2003.87	& 1.57$\pm$0.12		& 51.239	& 27.59	& VLA B (8)	\\
4.994	& 2003.98	& 1.55$\pm$0.01		& 51.246	& 27.51	& MERLIN (8)	\\
4.546	& 2004.14	& 1.80$\pm$0.23		& 51.246	& 27.51	& MERLIN (8)	\\
5.186	& 2004.14	& 1.48$\pm$0.18		& 51.246	& 27.51	& MERLIN (8)	\\
\hline
8.440	& 1989.92	& 0.65$\pm$0.28		& 51.291	& 26.93	& VLA B (9)	\\
8.460	& 2003.79	& 0.99$\pm$0.08		& 51.248	& 27.51	& VLA AB (10)	\\
8.435	& 2003.87	& 0.63$\pm$0.11		& 51.244	& 27.49	& VLA B (8)	\\
\hline
14.965	& 2003.87	& $\leq$0.25		& -		& -	& VLA B (8)	\\
\end{tabular}
\caption{\label{fluxtab}Flux densities measured from archival and new data.  References - (1) Hummel et al 
1985, (2) Vila et al 1990, (3) Mulder et al 1995, (4) FIRST, (5) Kregel \& Sancisi 2001 (note 
that the flux density of this point has been corrected, see section 3.1), (6) MERLIN archive, (7) MERLIN 
archive, (8) This Paper, (9) VLA archive, (10) Van Dyk, private communication.}
\end{table*}

\section{Results}

\subsection{Flux density}

The source is clearly visible in the 5 and 8.4\,GHz 2003 VLA images of NGC\,3310 and 
was initially assumed to be an extragalactic radio source, largely 
on the grounds that it is well outside the NGC\,3310 starburst. This 
conclusion was also reached by Kregel \& Sancisi (2001) who gave its 
flux density as 5.5\,mJy based on WSRT 21\,cm continuum observations. 
However the 1.4\,GHz  image of NGC\,3310 observed in 1982 by Hummel et al 
(1985), with an rms noise of 0.23\,mJy\,beam$^{-1}$, showed little evidence for a 
radio source at this position with an upper limit $<$1\,mJy.  In addition, 
the 1987 21\,cm continuum image of Mulder et al (1995) also shows no 
evidence for a radio source $>$1\,mJy.

The flux density of 5.5\,mJy at 1.4\,GHz, published by Kregal \& Sancisi (2001), includes data from Mulder et 
al (1995) which comprise 30\,per\,cent of the total.  Assuming these early data refer to the pre-supernova 
state, we estimate the 1997 1.4\,GHz flux density of the source to be $\sim7$\,mJy and this estimate has been 
included in Table~\ref{fluxtab}.

\begin{figure}
\psfig{figure=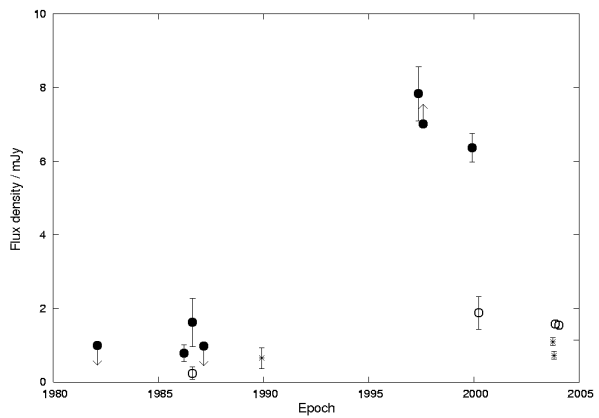,width=8cm}
\caption{\label{allfluxes}All measured flux densities.  Filled circles, open circles and 
crosses indicate 1.4, 5 and 8.4\,GHz points respectively.  Note the later 1997 point has been 
corrected, see section 3.1.}
\end{figure}

Fig.~\ref{allfluxes} shows the flux densities presented in 
Table~\ref{fluxtab}. This plot illustrates a large increase in 
1.4\,GHz luminosity (at least 5-fold) between data observed before and 
after the early 1990s.

From the FIRST and MERLIN measurements, the 1.4\,GHz flux density appears 
to have decreased between 1997 and 2000 by $\sim$10\,per\,cent per year, even 
though there is a large difference in beam size.  A similar decrease 
($\sim$10\,per\,cent per year) may also be present in the 5\,GHz flux density 
measurements between 2000 and 2003, however the data are not well sampled and further 
measurements, particularly at 1.4\,GHz, are required to confirm the amplitude 
of this decrease.

Much of the early data give only upper limits, particularly as the weak
source becomes confused with the extended galaxy emission in the lower
angular resolution measurements.  However, it can be seen from the early higher 
resolution data that a weak source was present within an arc second of the current 
position of \src with a flux density at the $\sim$0.5\,mJy level.

\subsection {Spectral Index}

As can be seen from Table~\ref{fluxtab}, the flux densities have been measured 
at several frequencies, with different beams and mostly at different epochs, 
hence spectral indices can only be  estimated. Nevertheless, it is clear from 
Table~\ref{fluxtab} that the source has a relatively steep spectrum.  MERLIN
measurements at 1.6 and 5\,GHz, measured in late 1999 and early 2000 
respectively, indicate a spectral index, $\alpha$ (where $S\propto\nu^{+\alpha}$), 
of $-0.95 \pm 0.3$ while the spectral index measured between 5\,GHz and 8.4\,GHz 
with the VLA in 2003 appears to be even steeper ($-1.67 \pm 0.49$). In order to check this 
result further, MERLIN measurements (see Table 1) were made in January 2004 at 
4546, 4994 and 5186 MHz, and from these observations $\alpha = -1.49 \pm 0.26$. 
These measurements are clearly all consistent with a steep spectral index, 
although a more accurate determination will require further simultaneous 
measurements at more widely spaced frequencies.

\subsection {Angular size}

The archival MERLIN observations at 1.6 and 5\,GHz show the source to be 
unresolved by the 150\,mas and 50\,mas beams. The 4.994\,GHz data-sets taken in 
December 2003 and January 2004 have been combined into a single high signal to 
noise image and uniform weighting of the data has been used to give
an equivalent beam of 35\,mas. There is no evidence for any extended structure and 
Gaussian fitting to the resulting cleaned image (Fig.~\ref{merlin}) gives an 
upper limit of $<$17\,mas (1.1\,pc) on the size of the source.

\section {Discussion}

\subsection {A radio supernova?}

The source is clearly not in the central starburst region of NGC\,3310, 
however it does appear to be associated with a cluster of H{\sc ii} regions 
(Hodge \& Kennicutt 1983) in the outer part of NGC\,3310 (Fig.~\ref{imagec}). 
Hence there are young massive stars  present which will 
result in supernovae at some stage.  The rise, and possible decay, of the 
radio light curve (Fig.~\ref{allfluxes}) is strong evidence in favour of 
this hypothesis.  Although these data are sparsely sampled, there is no 
doubt that at least a fivefold increase in flux density took place sometime 
in the early half of the 1990s.

If the source is at the distance of NGC\,3310, then its 5\,GHz luminosity is 
$\sim3\times10^{19}$\,W\,Hz$^{-1}$ which is comparable to the peak 
radio luminosity of many of the type II radio supernovae listed, 
for example, by Weiler et al 2002 (hereafter W02).

Since its rise in the  1990s, the radio flux density has remained high
until 2003, although there is some evidence that the flux density may be
decreasing at $\sim10$\,per\,cent per year since 1997.  Hence significant radio
emission has persisted for at least 5000 days.  It is well established (e.g.
W02) that a number of type II radio supernovae have radio
emission lasting for several thousand days.  The best example of this is
SN1979C (Montes et al 2000) in which the radio luminosity at 1.4 and 5\,GHz
has decreased by less than a factor of 3 over two decades.
This timescale is much
longer than the comparable timescale ($<$100 days) for type Ib/c
supernovae (e.g. W02) and hence would appear to rule out
identifying \src with this type of radio supernova.

From the observations we can only be sure that the strong radio emission from 
\src  appeared within a 7.5 year period between December 1989 and May 1997, however by 
considering the subsequent decrease in flux density we can put crude constraints on the age 
of the radio supernova.
W02 note that the decay parameter $\beta$ (which relates flux 
density, $S$, and time, $t$ by $S\,\propto\,t^{\beta}$) in models of known supernovae 
ranges from $-0.7$ to $-1.87$ for type II, and from $-1.24$ to $-1.57$ for type Ib/c radio 
supernovae.  If \src is assumed to be a supernova which exploded in 1990, 
then the 5\,GHz flux densities measured in  2000 (MERLIN) and 2003 (VLA) give 
a value for $\beta$ of $\sim -0.6$. A similar value ($\beta \sim -0.8$) is 
obtained from 1.4\,GHz measurements between 1997 (VLA) and 1999 (MERLIN).
These values of $\beta$ are low, although just consistent within the measurement 
errors with type II supernovae. However if we assume  that the supernova exploded 
in 1995, $\beta$ becomes $\sim -0.3$ at both frequencies, suggesting
that the supernova event took place early in the 7.5 year period.

The spectral index of \src is not well determined, but initial estimates suggest 
it to be significantly steeper than the spectral indices of typical type II radio supernovae 
listed by W02.  Further observations, using several frequencies, similar angular 
resolutions and similar epochs, are planned in order to determine the spectral index of \src 
with higher precision. 

The upper limits to the angular size of the source imply a linear size of
less than 1.1\,pc if it is at the distance of NGC\,3310.  Hence if the source is a 
radio supernova which exploded around 1990, this would imply an upper limit 
to the radial expansion velocity of $5 \times 10^4 $ km\,s$^{-1}$.
Assuming a typical expansion rate of 10$^4$\,km\,s$^{-1}$ (e.g. Pedlar et al 1999) then at the 
distance of NGC\,3310 a 15 year old supernova remnant would have an angular size of $\sim$2\,mas.
Global VLBI observations have now been proposed in order to investigate the structure on mas scales. 

The weak pre-1997 emission is within an arc second of the position of \src$\!\!$.
The spectral index, determined from the 1986 data, is 
$\alpha^{1.49}\!\!\!\!\!\!\!\!\!_{4.86} \sim -1.6 \pm 1.3$.  
This emission would not be expected to be directly associated with the radio
supernova itself.  Instead, it may be associated with either H{\sc ii}
regions or older supernova remnants.

\subsubsection { Evidence from other wavebands}

A search by Roberts \& Warwick (2000) of $ROSAT$ archive data observed 
between 1990 and 1997,  reported a detection of an X-ray source (NGC\,3310 X-2) with a 
luminosity of L $= 1.78\times10^{38}$\,erg\,s$^{-1}$ to the SE of NGC\,3310, and  \src 
lies within its 95\,per\,cent uncertainty ellipse (a radius of 15\,arcsec). 
Its  X-ray luminosity is consistent with the emission from other radio supernovae which 
show typical X-ray luminosities ranging from $10^{37}$ to $10^{39}$\,ergs\,s$^{-1}$ 
(e.g. Gallimore \& Beswick 2004 and references therein).

A search of the NASA extragalactic database (NED) for ground based optical 
observations of NGC\,3310, particularly during the early 1990s, reveals no obvious optical 
counterpart at the position of the radio supernova.  A number of archival $HST$ observations of NGC\,3310 
have also been examined but unfortunately do not include the position of the source.  We are continuing to 
search for relevant optical data but as the date of the supernova explosion is not known more 
accurately than to within a few years, it is uncertain whether any evidence will be found.

It may be that the supernova was embedded in a dense molecular cloud and hence any optical emission 
may have been heavily reddened. It is interesting to note that no optical counterpart was initially 
noticed corresponding to the detection of a radio supernova in NGC\,4258  by van der Hulst et al 
(1983).  A retrospective search of optical images showed a weak 13th magnitude optical source, 
heavily reddened by dust.

\subsection {Other possibilities}

There remains a small possibility that \src may be either a compact steep spectrum (CSS) or a GHz peaked 
spectrum (GPS) source, however these sources are not usually as variable as \src (O'Dea 1998).  In a 
study conducted by Seielstad, Pearson \& Readhead (1983), 90\,per\,cent of the steep spectrum sources in 
their sample do not vary significantly over the 4 years of their observations.  Also, these 
classes of object are often doubles or triples when imaged at the $\sim$50\,mas resolution of 
MERLIN.  The Global VLBI observations which have been proposed in order to resolve the radio supernova will also 
be a decisive method of ruling out, or otherwise, the compact structure expected in CSS or GPS 
sources.

A further possibility is that the source is a foreground pulsar, the source variations being due to 
scintillation.  Typically, pulsars have $\alpha = -1.6$ (e.g. Lorimer et al 1995). 
\src$\!\!$, however, is not in existing pulsar catalogues (M. Kramer - Private Communication) 
despite its relatively high integrated flux density.  Also the recent 5\,GHz MERLIN observations show 
that, unlike most pulsars, the source  is  unpolarised.  Furthermore the 5\,GHz data observations
carried out between December 2003 and January 2004 show little evidence for  the short timescale 
scintillation variability expected in a pulsar at the Galactic latitude ($b=54^o$) of \src$\!\!$.

\section{Conclusions}

The radio properties  of the compact, steep spectrum source $\sim$4\,kpc from the centre of 
NGC\,3310, appear to be similar to those of known radio supernovae. Both the light curve and the radio 
luminosity suggest that the source is a type II radio supernova, rather than a type Ib/c (W02).  
However, the radio spectral index appears to be steeper than typical radio supernovae, 
although further observations are required to confirm this.  This source will continue to be included 
in our ongoing monitoring program and Global VLBI observations have been applied for, which, assuming 
reasonable expansion velocities, should resolve the radio structure of the radio supernova.  
Retrospective optical observations are also required to help confirm or 
refute these conclusions.

\subsection*{Acknowledgements}

We thank Michael Kramer for useful discussions and the referee Schuyler Van Dyk for providing 
additional data.
The National Radio Astronomy Observatory is a facility of the National
Science Foundation operated under cooperative agreement by Associated Universities, Inc.
MERLIN is run by the University of Manchester as a National Facility on behalf of PPARC.
This research has made use of the NASA/IPAC Extragalactic Database (NED) which is operated by the Jet 
Propulsion Laboratory, California Institute of Technology, under contract with the National Aeronautics and 
Space Administration.
The Digitised Sky Survey was produced at the Space Telescope Science Institute under U.S. Government grant 
NAG W-2166.
MKA \& MS acknowledge  support from PPARC studentships.
RJB acknowledges PPARC support.

{}

\end{document}